\documentclass[twocolumn]{revtex4}%
\usepackage{amsmath}
\usepackage{graphicx}
\usepackage{amsfonts}
\usepackage{amssymb}
\begin{document}
\title{Ray-based description of normal mode amplitudes in a range-dependent waveguide }
\author{A.L. Virovlyansky, A.Yu. Kazarova, L.Ya. Lyubavin}
\affiliation{Institute of Applied Physics, 46 Ulyanova St., Nizhny
Novgorod, 603950 Russia } 

\begin{abstract}
An analogue of the geometrical optics for description of the modal structure
of a wave field in a range-dependent waveguide is considered. In the scope of
this approach the mode amplitude is expressed through solutions of the ray
equations. This analytical description accounts for mode coupling and remains
valid in a nonadiabatic environment. It has been used to investigate the
applicability condition of the adiabatic approximation. An applicability
criterion is formulated as a restriction on variations of the action variable
of the ray.


\end{abstract}
\maketitle

\section{Introduction \label{sec:intro}}

We consider a method for analysis of the modal structure of a scalar wave
field in a two dimensional range-dependent waveguide. This method derived in
Refs. \cite{V99b,V2001a,V2004} (see also \cite{BZ79}) is based on projecting
the ray representation of the wave field onto normal modes and evaluating the
corresponding integrals using the stationary phase technique. As a result the
mode amplitude is expressed through parameters of ray trajectories. This
approach establishes a relationship between the ray and mode representations
of the wave field in a range-dependent waveguide and provides an approximate
solution to the mode coupling equation.

In describing ray trajectories we apply the Hamiltonian formalism taken in
terms of the action-angle variables \cite{LLmech,AZ91}. It is shown that each
mode is formed by contributions from rays whose action variables up to a
multiplicative constant are equal to the mode number. In the nonadiabatic
environment the action varies along the ray path. Our objective in this work
is to find out how the nonadiabaticity of rays manifests itself in variations
of mode amplitudes. In the nonadiabatic waveguide an initially excited single
mode breaks up into a group of modes. We argue that the width of the group is
proportional to the spread of ray actions at the given range.

The formula for the mode amplitude, valid in the nonadiabatic waveguide, has
been used to formulate an applicability condition of the adiabatic
approximation. In contrast to traditional approaches for solving this problem
\cite{BL91,BG99} we proceed not from the mode coupling equation but from its
solution (albeit approximate). An applicability criterion discussed in the
present paper takes into account accumulation of errors with range.

All results are obtained for the wave field that obeys the parabolic equation.
This assumption is not necessary: the results are readily generalized to the
field governed by the Helmholtz equation. We use the parabolic equation
approximation for two reasons. First, unlike the Helmholtz equation it can be
easily solved numerically which is very important for testing our approximate
formulas. Besides, the parabolic equation formally coincides with the
Schr\"{o}dinger equation and therefore formulas derived in this work may be
applied in quantum mechanics to study the motion of a particle in a potential
well with time-dependent parameters.

In the present paper we investigate general properties of wave propagation in
a range-dependent waveguide \ by the example of the underwater sound channel.
This choice is caused by interests of the authors and by the fact that the
topics addressed here are of importance in underwater acoustics
\cite{BL91,BG99,Shang88,Godin2002}.

The paper is organized as follows. Section \ref{sec:ray-rep} provides a brief
description of the ray representation of the wave field. Main relations of the
Hamiltonian formalism expressed in terms of both position-momentum and
action-angles variables are presented. In Sec. \ref{sec:WKB-P} we consider the
mode representation (in the WKB approximation) of the wave field and the mode
coupling equation. We also outline a traditional method for estimating the
validity region of the adiabatic approximation by using the perturbation
theory for the mode coupling equation. In Sec. \ref{sec:projection} formulas
expressing mode amplitudes through parameters of ray trajectories are derived.
Section \ref{sec:variations} focuses on connection between variations of the
action variable and mode coupling. It is shown how a standard ray tracing can
be used to estimate the number of interacting modes. An applicability
criterion for the adiabatic approximation is formulated as restriction on the
spread of ray actions. These results are verified and illustrated using
numerical examples. The conclusions are summarized in the final section.

\section{Ray representation of the wave field \label{sec:ray-rep}}

\subsection{Parabolic equation approximation \label{sub:parabolic}}

Consider a monochromatic wave field at a carrier frequency $f$ in a
two-dimensional acoustic waveguide with the sound speed $c$ being a function
of depth, $z,$ and range, $r$. We shall assume that at each range point the
sound speed profile has a single minimum. The wave field $u(r,z)$ is governed
by the Helmholtz equation \cite{BL91,BG99,JKPS94}
\begin{equation}
\frac{1}{r}\frac{\partial}{\partial r}r\frac{\partial u}{\partial r}%
+\frac{\partial^{2}u}{\partial z^{2}}+\frac{\Omega^{2}}{c^{2}(r,z)}u=0,
\label{Helmholtz}%
\end{equation}
where $\Omega=2\pi f$. Select a reference sound speed $c_{0}$ such that
$\left|  c(r,z)-c_{0}\right|  <<c_{0}$ (in underwater acoustics such choice of
$c_{0}$ is always possible) and assume that grazing angles of the propagating
waves are small. Then it is convenient to introduce an envelope function
$v(r,z)$ connected to $u(r,z)$ by an expression%
\begin{equation}
u=\frac{v}{\sqrt{r}}e^{ikr}, \label{u-v}%
\end{equation}
with $k=\Omega/c_{0}$. This function is approximately described by the
standard parabolic equation \cite{BL91,JKPS94,SFW97}
\begin{equation}
2ik\frac{\partial v}{\partial r}+\frac{\partial^{2}v}{\partial z^{2}}%
-2k^{2}Uv=0, \label{parabolic}%
\end{equation}
where%
\begin{equation}
U(r,z)=\frac{1}{2}\left(  1-\frac{c_{0}^{2}}{c^{2}(r,z)}\right)  . \label{U}%
\end{equation}
Notice that Eq. (\ref{parabolic}) formally coincides with the time-dependent
Schr\"{o}dinger equation. In this analogy $r$, $k^{-1}$, and $U(r,z)$ play
roles of time, Planck's constant, and potential, respectively.

\subsection{Hamiltonian formalism in terms of momentum-position variables
\label{sub:p-z}}

In the geometrical optics approximation the wave field is formed by
contribution from all rays (eigenrays) arriving at the observation point. The
ray trajectory is defined by the Hamilton equations \cite{AZ91,SFW97}
\begin{equation}
\frac{dz}{dr}=\frac{\partial H}{\partial p},\;\;\;\frac{dp}{dr}=-\frac
{\partial H}{\partial z} \label{dzdr}%
\end{equation}
with the Hamiltonian
\begin{equation}
H=p^{2}/2+U. \label{H-P}%
\end{equation}
The momentum $p$ is connected to the ray grazing angle $\chi$ through the
relation%
\begin{equation}
p=dz/dr=\tan\chi. \label{p-chi}%
\end{equation}

A contribution to the wave field from an eigenray is presented in the form%
\begin{equation}
v=Ae^{ikS}, \label{u-geom-P}%
\end{equation}
where $S$ and $A$ are the ray eikonal and amplitude, respectively. The eikonal
$S$ is an analog to Hamilton's principal function in classical mechanics and
it is given by an integral%
\begin{equation}
S=\int\left(  pdz-Hdr\right)  \label{S}%
\end{equation}
running over the ray path \cite{LLmech,SFW97,BornWolf}. An explicit expression
for the ray amplitude depends on the source exciting the wave field. In the
case of a point source the wave field is determined by Eq. (\ref{parabolic})
with an initial condition
\begin{equation}
v(0,z)=\delta(z-z_{0}). \label{point-ini}%
\end{equation}
Then all rays escape the same point $(0,z_{0})$ and each trajectory can be
labelled by the ray starting momentum $p_{0}$. The amplitude
\begin{equation}
A=\sqrt{\frac{k}{2\pi i\left\vert \partial z/\partial p_{0}\right\vert }%
}e^{-i\mu\pi/2}, \label{pointP}%
\end{equation}
where $\mu$ is the Maslov index or integral number of times that the ray
passes through caustics \cite{Gutz67}.

\subsection{Action-angle variables \label{sub:ac-an-P}}

Since ray trajectories in the waveguide are oscillating curves, their analysis
may be simplified by using the so-called action-angle variables
\cite{LLmech,AZ91}. In mechanics the latter are often applied to study
oscillations of particles in potential wells.

\subsubsection{Range-independent waveguide \label{sub:range-indep}}

First, consider a range-independent waveguide where the Hamiltonian $H$
remains constant along the ray path (analog to the energy conservation law in
mechanics). The action variable $I$ is defined by the integral
\cite{LLmech,AZ91}\
\begin{equation}
I=\frac{1}{2\pi}\oint p\,dz=\frac{1}{\pi}\int_{z_{\min}}^{z_{\max}}%
dz\sqrt{2\left[  H-U(z)\right]  },\, \label{I-P}%
\end{equation}
where $z_{\min}$ and $z_{\max}$ are the lower and upper ray turning depths,
respectively, satisfying the condition $U(z)=H$. Equation (\ref{I-P}) defines
the \textquotedblleft energy\textquotedblright\ $H$ as a function of the
action $I$. The ray trajectory is a periodic curve whose period along the
$r$-axis (the cycle length), $D$, and the angular frequency of spatial
oscillations, $\omega$, are determined by the relation
\begin{equation}
\frac{dH}{dI}=\omega=\frac{2\pi}{D}. \label{omega}%
\end{equation}

Take one cycle of the unperturbed ray path that begins at the minimum of the
trajectory. At the first half-cycle the canonical transformation,
\begin{equation}
p=p(I,\theta),\quad z=z(I,\theta),\label{canon}%
\end{equation}
from the position-momentum, $(p,z)$, to the action-angle, $(I,\theta)$,
variables is determined by the equations \cite{LLmech,AZ91,Abdullaev}
\begin{equation}
p=\frac{\partial G}{\partial z},\quad\theta=\frac{\partial G}{\partial
I},\label{canon1}%
\end{equation}
with%
\begin{equation}
G(I,z)=\int_{z_{\min}}^{z}dz\sqrt{2\left[  H(I)-U(z)\right]  }\label{G-P}%
\end{equation}
being a generating function of the canonical transformation. At this
half-cycle the angle variable $\theta$ varies from $0$ to $\pi$. The
transformation is continued to the next half-cycle ($\pi<\theta<2\pi$) using
relations $p(I,\theta)=-p(I,2\pi-\theta)$, and $z(I,\theta)=z(I,2\pi-\theta)$.
The ray equations in the new variables take the trivial form
\begin{equation}
\frac{dI}{dr}=-\frac{\partial H}{\partial\theta}=0,\quad\frac{d\theta}%
{dr}=\frac{\partial H}{\partial I}=\omega(I).\label{dIdr}%
\end{equation}
Note, that the so defined angle variable $\theta$ varies from $0$ to $2\pi$ at
a part of the trajectory beginning at one minimum and ending at the next one.
To make the angle variable continuous, its value should be increased by $2\pi$
at the beginning of each new cycle. Both functions in Eq. (\ref{canon}) are
periodic in $\theta$ with period $2\pi$. Correspondingly, an arbitrary
function $F(z)$ expressed through the action-angle variables becomes a
periodic function of $\theta$ and can be expanded in a Fourier series%
\begin{equation}
F(z)=\sum_{\nu=0}^{\infty}F_{\nu}(I)\cos\nu\theta\text{.}\label{F-I-T}%
\end{equation}
The coefficients $F_{\nu}(I)$ are expressed analytically only for a few
special examples of $U(z)$ and $F(z)$. But numerical evaluation of these
coefficients can be easily performed using a standard ray code. If we denote
the ray trajectory with the action variable $I$ by $z(r)$, then%
\[
F_{\nu}(I)=\frac{q_{\nu}}{D(I)}%
\]
\begin{equation}
\times\int_{0}^{D(I)}dr~F(z(r_{\min}+r))\cos\left(  \nu\omega(I)(r_{\min
}+r)\right)  ,\label{F-nu}%
\end{equation}
where $q_{0}=1$, $q_{\nu}=2$ for $\nu>0$, and $r_{\min}$ is a range
corresponding to a minimum of the ray path.

\subsubsection{Range-dependent waveguide \label{sub:range-dep}}

In a range-dependent waveguide canonical transformations from $(p,z)$ to
$(I,\theta)$ variables are different at different range points. At each
particular range $r$ the transformation is defined using the so-called
reference waveguide: an imaginary \textbf{range-independent} waveguide whose
cross-section coincides with that of a real one at\ the given range. The
connection between pairs $(p,z)$ and $(I,\theta)$ at range $r$ is determined
by formulas present in Sec. \ref{sub:range-dep} that should be applied in a
corresponding reference waveguide. Since the functions $p(I,\theta)$,
$z(I,\theta)$, $G(I,z)$, $H(I)$, $\omega(I)$, and $F_{\nu}(I)$ become
different at different ranges they acquire an additional argument $r$ that
labels the reference waveguide.

The canonical transformation is determined by the equation \cite{LLmech}
\begin{equation}
dS=pdz-Hdr=dG-\theta dI-\bar{H}dr, \label{dS}%
\end{equation}
where $\bar{H}$ is a new Hamiltonian. According to this relation
\begin{equation}
\bar{H}(I,\theta,r)=H(I,r)+\Lambda(I,\theta,r), \label{Hbar}%
\end{equation}
with
\begin{equation}
\Lambda(I,\theta,r)=\left.  \frac{\partial G(I,z,r)}{\partial r}\right\vert
_{z=z(I,\theta,r)}. \label{LambdaH}%
\end{equation}

The Hamilton equations retains their canonical form%
\begin{equation}
\frac{dI}{dr}=-\frac{\partial\bar{H}}{\partial\theta}=-\frac{\partial\Lambda
}{\partial\theta},\;\frac{d\theta}{dr}=\frac{\partial\bar{H}}{\partial
I}=\omega+\frac{\partial\Lambda}{\partial I},\label{dIdr-H-1}%
\end{equation}
where%
\begin{equation}
\omega(I,r)=\frac{\partial H(I,r)}{\partial I}.\label{om-H-1}%
\end{equation}
The ray eikonal in the new variables can be found by integrating Eq.
(\ref{dS}). After some algebra we get%
\[
S=\int_{0}^{r}\left(  Id\theta-\left(  H+\Lambda\right)  dr\right)
\]%
\[
-G(z(0),I(0),0)+G(z(r),I(r),r)
\]%
\begin{equation}
+\theta(0)I(0)-\theta(r)I(r).\label{S-I-T}%
\end{equation}
Initial and final values of the angle variable, $\theta(0)$ and $\theta(r)$,
present in the right hand side should be taken modulo $2\pi$.

\bigskip The term $\Lambda$ can be expressed through coefficients of an
expansion
\begin{equation}
\left.  \frac{\partial U(r,z)}{\partial r}\right\vert _{z=z(I,\theta,r)}%
=\sum_{\nu=0}^{\infty}V_{\nu}(I,r)\cos\theta\label{dU-IT}%
\end{equation}
analogous to Eq. (\ref{F-I-T}). First, notice that%
\[
\frac{\partial}{\partial r}G(I,z,r)=\pm\int_{z_{\min}}^{z}\left(
\frac{\partial H(I,r)}{\partial r}-\frac{\partial U(r,z)}{\partial r}\right)
\]

\begin{equation}
\times\frac{dz}{\sqrt{2\left[  H(I,r)-U(r,z)\right]  }},\label{dGdI}%
\end{equation}
where $\pm$ denotes the sign of $p$ at a current half-cycle. From Eq.
(\ref{I-P}) it follows that%
\[
\frac{\partial H(I,r)}{\partial r}=\frac{\omega(I,r)}{\pi}%
\]%
\begin{equation}
\times\int_{z_{\min}}^{z_{\max}}\frac{\partial U(r,z)}{\partial r}\frac
{dz}{\sqrt{2\left[  H(I,r)-U(r,z)\right]  }}.\label{dHdr}%
\end{equation}
The right hand side, formally, can be considered as an integral over a
half-cycle of the ray path in the reference waveguide. Then
\[
\frac{\omega(I,r)dz}{\sqrt{2\left[  H(I,r)-U(r,z)\right]  }}=d\theta
\]
and%
\begin{equation}
\frac{\partial H(I,r)}{\partial r}=V_{0}(I,r).\label{dHdr-I-T}%
\end{equation}
Similarly it can be shown that%
\[
\int_{z_{\min}}^{z}\frac{\partial U(r,z)}{\partial r}\frac{dz}{\sqrt{2\left[
H(I,r)-U(r,z)\right]  }}%
\]%
\begin{equation}
=\frac{1}{\omega}\left(  \theta V_{0}(I,r)+\sum_{\nu=1}^{\infty}V_{\nu
}(I,r)\frac{\sin(\nu\theta)}{\nu}\right)  .\label{dU-I-T}%
\end{equation}

Combining Eqs. (\ref{dGdI}), (\ref{dHdr-I-T}), and (\ref{dU-I-T}) we arrive at%
\begin{equation}
\Lambda=-\frac{1}{\omega}\sum_{\nu=1}^{\infty}V_{\nu}(I,r)\frac{\sin\left(
\nu\theta\right)  }{\nu}. \label{dGdr}%
\end{equation}
Note that the coefficients $V_{\nu}(I,r)$ can be calculated using Eq.
(\ref{F-nu}) in a reference waveguide corresponding to range $r$ with $F$
replaced by $V(r,z)=\partial U(r,z)/\partial r$. When evaluating the right
hand side of Eq. (\ref{F-nu}) one should integrate over the ray path computed
in the reference waveguide but not in the real one. In this calculation the
argument $r$ of function $V(r,z)$ must be considered as a constant labelling
the reference waveguide.

The adiabatic approximation for rays is valid if the sound speed $c$ is a so
slow function of range that $\Lambda$ becomes negligible. Then the action
variable $I$ remains constant along the ray path \cite{LLmech,BL91}.

\section{Mode representation in the WKB approximation \label{sec:WKB-P}}

\subsection{Eigenfunctions and eigenvalues \label{subsub:eigen-P}}

The normal mode representation of the wave field in a range-independent
waveguide is given by an expansion into a sum of eigenfunctions of the
Sturm-Liouville eigenvalue problem \cite{BG99,LLquant}%
\begin{equation}
\frac{1}{2}\frac{d^{2}\varphi_{m}}{dz^{2}}+k^{2}(H_{m}-U)\varphi_{m}=0
\label{eigen-phi}%
\end{equation}
with appropriate boundary conditions. The latter can be determined by
reflection coefficients at the surface, $V_{s}=e^{i\phi_{s}}$, and at the
bottom, $V_{b}=e^{i\phi_{b}}$. We assume that reflections at boundaries occur
without energy loss and therefore $\phi_{s}$ and $\phi_{b}$ are real
constants. In particular, for a pressure release surface and rigid bottom we
have $\phi_{s}=\pi$ and $\phi_{b}=0$. In case when the upper (lower) mode
turning points lie within a water bulk we have $\phi_{s}=-\pi/2$ ($\phi
_{b}=-\pi/2$). The eigenfunctions are orthogonal and normalized in such a way that%

\begin{equation}
\int dz~\varphi_{m}\varphi_{n}=\delta_{mn}. \label{ort}%
\end{equation}
In the WKB approximation the eigenvalue of the $m$-th mode is
\begin{equation}
H_{m}=H(I_{m}), \label{Hm-P}%
\end{equation}
where $I_{m}$ satisfies the quantization rule \cite{BL91,LLquant}%
\begin{equation}
kI_{m}=m-\frac{\phi_{s}+\phi_{b}}{2\pi}. \label{I-m}%
\end{equation}
The quantity $I_{m}$ can be treated as an action variable associated with the
$m$-th mode: if only this mode is excited then all rays have the same action
$I=I_{m}$. Using Eq. (\ref{omega}) we get a simple approximate relation for
the difference between eigenvalues of neighboring modes \cite{BL91}
\begin{equation}
\frac{dH_{m}}{dm}=H_{m+1}-H_{m}=\omega_{m}/k, \label{dkm-dm}%
\end{equation}
where $\omega_{m}\equiv\omega(I_{m})$.

The $m$-th eigenfunction $\varphi_{m}(z)$ between its turning points can be
represented as \cite{LLquant}%

\begin{equation}
\varphi_{m}(z)=\varphi_{m}^{+}(z)+\varphi_{m}^{-}(z), \label{phi-WKB}%
\end{equation}
where%

\begin{equation}
\varphi_{m}^{\pm}(z)=Q_{m}e^{\pm i\left(  kS_{m}(z)+\phi_{b}/2)\right)
},\label{phipm}%
\end{equation}%
\begin{equation}
\;S_{m}(z)=\int_{z_{\min}}^{z}dz\,p_{m}(z),\;\;p_{m}(z)=\sqrt{2\left[
H_{m}-U(z)\right]  },\label{Sm}%
\end{equation}
and%
\begin{equation}
Q_{m}=\sqrt{\frac{\omega_{m}}{2\pi p_{m}(z)}}.\;\label{Q-m-P}%
\end{equation}

\subsection{Mode coupling equation \label{sub:mode-coupl}}

\bigskip In each cross-section of a waveguide the field can be decomposed into
a sum of the local modes, that is the modes of the reference waveguide (see
Sec. \ref{sub:range-dep}) corresponding to this particular range. Since at
different range points we deal with, generally, different reference
waveguides, the eigenfunctions and eigenvalues become functions of range. The
mode representation of the wave field in a range-dependent waveguide has the
form%
\begin{equation}
u(r,z)=\sum_{m}a_{m}(r)\varphi_{m}(r,z), \label{modrep}%
\end{equation}
where we have emphasized the $r$ dependence of the eigenfunction. By
substituting this expression in Eq. (\ref{parabolic}) and using the
orthogonality condition (\ref{ort}) we get the mode coupling equation
\begin{equation}
\frac{da_{m}}{dr}+ikH_{m}a_{m}=-\sum_{m_{1}}a_{m_{1}}\int dz~\frac
{\partial\varphi_{m_{1}}}{\partial r}\varphi_{m}. \label{coupl-P}%
\end{equation}
For a point source defined by Eq. (\ref{point-ini}) it should be solved with
an initial condition%
\begin{equation}
a_{m}(0)=\varphi_{m}(0,z_{0}). \label{am-point-0}%
\end{equation}
For later convenience we present the derivative $\partial\varphi_{m}/\partial
r$ as an expansion
\begin{equation}
\frac{\partial\varphi_{m}}{\partial r}=\sum_{\nu\neq0}B_{m\nu}\varphi_{m+\nu
}.\; \label{phi-r}%
\end{equation}
Substituting this in Eq. (\ref{coupl-P}) yields
\begin{equation}
\frac{da_{m}}{dr}+ikH_{m}a_{m}=\sum_{\nu\neq0}B_{m\nu}a_{m+\nu}.
\label{coupling-P}%
\end{equation}
It can be shown that
\begin{equation}
B_{m\nu}=\frac{1}{H_{m}-H_{m+\nu}}\int dz~\varphi_{m}\frac{\partial
U}{\partial r}\varphi_{m+\nu}. \label{Bmn0-P}%
\end{equation}
This formula is derived by differentiating Eq. (\ref{eigen-phi}) for the
reference waveguide with respect to $r$ and exploiting the normalization
condition (\ref{ort}). A detailed derivation of a more general result for the
wave field governed by the Helmholtz equation see in Ref. \cite{BG99}.

\subsubsection{Matrix elements\label{sub:matrix}}

In the WKB approximation the matrix element of a smooth function $F(z)$
\cite{LLquant}%
\begin{equation}
F_{m\nu}=\int dz~\varphi_{m}(z)F(z)\varphi_{\nu}(z)\label{Fmn0}%
\end{equation}
with $m,n>>1$ can be expressed through coefficients of expansion
(\ref{F-I-T}). Indeed, take $F(z)$ whose characteristic scale is much greater
than the wavelength $2\pi/k$. Then the matrix elements are very small unless
$\left\vert m-\nu\right\vert <<m$. Making use of Eqs. (\ref{phi-WKB}%
)-(\ref{Sm}) rewrite Eq. (\ref{Fmn0}) in the form%
\[
F_{m,m+\Delta m}=2\int_{z_{\min}}^{z_{\max}}dz~F(z)Q_{m}^{2}(z)
\]
\begin{equation}
\times\cos\left[  k\left(  S_{m+\Delta m}(z)-S_{m}(z)\right)  \right]
.\label{Fmn1}%
\end{equation}
From Eqs. (\ref{dkm-dm}) and (\ref{Sm}) it follows that%
\begin{equation}
k\left(  S_{m+\Delta m}(z)-S_{m}(z)\right)  =\omega_{m}\Delta m\int_{z_{\min}%
}^{z}\frac{dz}{p_{m}(z)}.\label{dSdm}%
\end{equation}
According to Eq. (\ref{p-chi}) $dz/p_{m}=dr$ and the integral on the right can
be considered as a shift along the $r$-axis between a minimum and a current
point of a ray trajectory with action $I_{m}$. In the range-independent
waveguide $\omega_{m}r=\theta$ and we get a desired connection between the
matrix elements and the coefficients of Eq. (\ref{F-I-T})%
\[
F_{m,m+\Delta m}=\frac{1}{2\pi}\int_{0}^{2\pi}d\theta~F(z(I_{m},\theta
))\cos\left(  \Delta m~\theta\right)
\]%
\begin{equation}
=\frac{1}{2}F_{\left\vert \Delta m\right\vert }(I_{m}).\label{matr-elem}%
\end{equation}
According to this result Eq. (\ref{Bmn0-P}) in the high frequency
approximation reduces to
\begin{equation}
B_{m\nu}=-\frac{k}{2\nu\omega_{m}}V_{\left\vert \nu\right\vert }%
(I_{m},r).\label{Bmn-P}%
\end{equation}

\subsection{\bigskip Adiabatic approximation for mode amplitudes
\label{sub:adiabatic-mode}}

In the adiabatic approximation the right hand side of the mode coupling
equation (\ref{coupling-P}) is assumed to be negligible \cite{BL91,BG99} and
we get%
\begin{equation}
a_{m}(r)=a_{m}(0)e^{-ik\int_{0}^{r}H_{m}(r^{\prime})dr^{\prime}}%
.\label{a-adiab-P}%
\end{equation}
The first order correction can be obtained using a simple perturbation theory.
Consider the case when only one mode with number $m_{0}$ is excited at $r=0$,
that is
\begin{equation}
a_{m}(0)=\delta_{mm_{0}}\text{.}\label{am0-P}%
\end{equation}
Substituting Eq. (\ref{a-adiab-P}) with $a_{m}(0)$ defined by Eq.
(\ref{am0-P}) into the right hand side of Eq. (\ref{coupling-P}) we find that
for $m\neq m_{0}$
\begin{equation}
a_{m}(r)=e^{-ik\int_{0}^{r}H_{m_{0}}(r^{\prime})dr^{\prime}}q_{m}%
,\label{am-qm-P}%
\end{equation}
where%
\begin{equation}
q_{m}=\int_{0}^{r}dr^{\prime}B_{m,m_{0}-m}(r^{\prime})e^{i(m-m_{0})\int
_{0}^{r^{\prime}}\omega_{m_{0}}(r^{\prime\prime})dr^{\prime\prime}%
}.\label{qm0}%
\end{equation}
At high frequencies, when Eq. ( \ref{Bmn-P}) is valid, $q_{m}$ can be
presented in the form%
\[
q_{m}=-\frac{k}{2(m-m_{0})}%
\]%
\begin{equation}
\times\int_{0}^{r}\frac{dr^{\prime}}{\omega_{m}(r^{\prime})}V_{\left\vert
m-m_{0}\right\vert }(r^{\prime},I_{m})e^{i(m-m_{0})\int_{0}^{r^{\prime}}%
\omega_{m_{0}}(r^{\prime\prime})dr^{\prime\prime}}.\label{qm-P}%
\end{equation}
The condition
\begin{equation}
\left\vert q_{m}\right\vert <<1\label{qm-small}%
\end{equation}
(or its analog for the Helmholtz equation) is traditionally considered as a
starting point for studying the applicability of adiabatic approximation
\cite{BL91,BG99}. In particular it turns out that the inequality
(\ref{qm-small}) requires that%
\begin{equation}
D/L<<1,\label{DL}%
\end{equation}
where $L$ is the characteristic scale of the horizontal range dependence.
Another criterion involving the dependence on a frequency of the propagating
wave has the form \cite{BG99,Milder69}%
\begin{equation}
kD^{2}/L<<1.\label{k2DL}%
\end{equation}
There are less general (but less restrictive and therefore more useful)
criteria that can be obtained for waveguides with some special properties
\cite{BG99}.

\section{Projection of the ray representation onto normal modes
\label{sec:projection}}

\subsection{Mode amplitude as a function of ray parameters
\label{sub:mode-ampl}}

In this section we show that the mode amplitude in a range-dependent waveguide
can be approximately expressed through parameters of ray trajectories. In more
details this issue has been considered in Refs. \cite{V99b,V2001a}. Our
starting point is an equation%
\begin{equation}
a_{m}(r)=\int dz~v(r,z)\varphi_{m}(r,z)\label{am-v}%
\end{equation}
that follows from the mode orthogonality condition (\ref{ort}). Substituting
$v(r,z)$ from Eq. (\ref{u-geom-P}) and using the WKB approximation for
$\varphi_{m}(r,z)$ we get $a_{m}=a_{m}^{+}+a_{m}^{-}$ with%
\begin{equation}
a_{m}^{\sigma}=\int dz~B(r,z,\sigma)e^{ik\Phi(r,z,\sigma)},\label{a-m-s}%
\end{equation}
where $\sigma$ denotes plus or minus,
\begin{equation}
B=e^{-i\mu\pi/2+i\sigma\phi_{b}/2}~AQ_{m},\label{B}%
\end{equation}
and%
\begin{equation}
\Phi=S+\sigma S_{m}.\label{Phi}%
\end{equation}
In the high frequency approximation (large $k$) integral (\ref{a-m-s}) can be
evaluated using a standard stationary phase technique \cite{BornWolf}. \ This
yields%
\[
a_{m}^{\sigma}=\left.  \sqrt{\frac{2\pi}{k\left\vert \partial^{2}\Phi/\partial
z^{2}\right\vert }}\right\vert _{z=z_{st}}%
\]%
\begin{equation}
\left.  B\exp\left[  ik\Phi+i\frac{\pi}{4}\text{sgn}\left(  \partial^{2}%
\Phi/\partial z^{2}\right)  \right]  \right\vert _{z=z_{st}},\label{technique}%
\end{equation}
with $z_{st}$ being a stationary phase point where the derivative
$\partial\Phi/\partial z$ vanishes. The function sgn($x$) gives the sign of
its argument. In classical mechanics it is well-known that the eikonal $S$
considered as a function of range $r$, starting coordinate $z_{0}\,$, and a
final coordinate $z$ obeys the relation \cite{LLmech}
\begin{equation}
\frac{\partial S}{\partial z}=p,\label{dSdz}%
\end{equation}
where $p$ is a momentum at range $r$. Combining this with Eq. (\ref{Sm}) we
get
\begin{equation}
\frac{\partial\Phi}{\partial z}=p+\sigma p_{m}.\label{dPhidz}%
\end{equation}
This derivative vanishes at a point $z$ such that a ray arriving at this point
has an action%
\begin{equation}
I=I_{m}\label{stationary}%
\end{equation}
and
\begin{equation}
\sigma=-\text{sgn}(p).\label{sigma}%
\end{equation}
The condition (\ref{stationary}) singles out rays whose actions at the given
range $r$ are equal to the action of the $m$-th mode. In quantum theory a
similar result was obtained in Ref. \cite{BZ79}.

For the second derivative of $\Phi$ we have an expression%
\[
\frac{\partial^{2}\Phi}{\partial z^{2}}=\frac{\partial p}{\partial z}%
-\sigma\frac{\partial U/\partial z}{p_{m}}%
\]%
\begin{equation}
=\frac{1}{p}\frac{\partial H}{\partial z}=\frac{\omega}{p}\frac{\partial
I}{\partial z},\label{d2Phi}%
\end{equation}
where $H$ and $I$ are considered as functions of $\ $the ray arrival depth at
the given range $r$. From Eq. (\ref{technique}) it follows that%
\[
a_{m}^{\sigma}=\left.  A\left\vert k\partial I/\partial z\right\vert
^{-1/2}\right\vert _{z=z_{st}}%
\]%
\begin{equation}
\left.  \exp\left(  ik\Phi-i\pi/4\sigma\text{sgn}(\partial I/\partial
z)\right)  \right\vert _{z=z_{st}}.\label{technique1}%
\end{equation}
To complete the calculation of the mode amplitude we should express the
derivative $\partial I/\partial z$ through parameters of rays. In the next
subsection this is done for two important examples of the starting field.

\subsection{Point source and source exciting a single mode
\label{sub:two-sources}}

In the case of a point source forming the starting field (\ref{point-ini}) we
deal with a congruence of rays leaving point $(0,z_{0})$. Labelling each ray
by its starting momentum $p_{0}$ we present the derivative $\partial
I/\partial z$ as $(\partial I/\partial p_{0})/(\partial z/\partial p_{0})$.
Then insertion of Eq. (\ref{pointP}) into Eq. (\ref{technique1}) yields%
\[
a_{m}^{\sigma}=\frac{1}{\sqrt{2\pi\left\vert \partial I/\partial
p_{0}\right\vert }}%
\]%
\begin{equation}
\times\exp\left(  k\left(  S+\sigma S_{m}\right)  -i\mu\pi/2+i\sigma\phi
_{b}/2+i(\beta-1)\pi/4\right)  ,\label{a-s-I}%
\end{equation}
with
\begin{equation}
\beta\equiv-\sigma\text{sgn}(\partial I/\partial p_{0})\text{sgn}(\partial
z/\partial p_{0}).\label{beta-P}%
\end{equation}
Formula (\ref{a-s-I}) accounts for a contribution from a ray satisfying the
condition (\ref{stationary}) to the $m$-th mode. The mode amplitude is
evaluated by summing up contributions from all such rays. Note that in the
range-independent waveguide at any range point and for any mode there are
exactly two rays satisfying Eq. (\ref{stationary})
\cite{V99b,V2001a,V2004,V97}. They escape the source at grazing angles (equal
in absolute value and opposite in sign) that coincides with grazing angles of
the quasi-plane waves $\varphi_{m}^{+}e^{-ikH_{m}r}$ and $\varphi_{m}%
^{-}e^{-ikH_{m}r}$ at depth $z_{0}$. This statement remains valid in an
adiabatic waveguide as well. Notice that if rays are adiabatic, i.e. the term
$\Lambda$ in the Hamiltonian (\ref{Hbar}) is negligible, then Eq.
(\ref{a-s-I}) reduces to Eq. (\ref{a-adiab-P}) with $a_{m}(0)$ defined by Eq.
(\ref{am-point-0}). A corresponding transformation is rather simple but
somewhat lengthy and therefore we do not present it here.

A similar result can be derived for a source exciting a single mode with
$m=m_{0}$ \cite{V99b}. In this case we have two congruences of rays associated
with functions $\varphi_{m_{0}}^{\pm}$. Initial actions of all rays are equal
to $I_{m_{0}}$ and their starting depths are located between turning points of
the mode. There are two rays escaping each point $z_{0}$ within this interval
of depths. These rays have starting momenta
\begin{equation}
p_{0}=\pm\sqrt{2(H_{m_{0}}(0)-U(z_{0}))}\label{p0}%
\end{equation}
and complex amplitudes \cite{V99b,Maslov}
\begin{equation}
A=\frac{Q_{m_{0}}(z_{0})}{\sqrt{\left\vert \partial z/\partial z_{0}%
\right\vert }}e^{\pm i\left(  S_{m_{0}}(z)+\phi_{b}/2\right)  }.\label{A-m0}%
\end{equation}
Here we label rays by their starting depths. Substituting Eq. (\ref{A-m0})
into Eq. (\ref{technique1}) and representing $\partial I/\partial z$ as
$(\partial I/\partial z_{0})/(\partial z/\partial z_{0})$ yields%
\[
a_{m}^{\sigma}=\frac{Q_{m_{0}}(z_{0})}{\sqrt{k\left\vert \partial I/\partial
z_{0}\right\vert }}%
\]%
\[
\times\exp\left(  ik\left(  S+\sigma S_{m}+\beta_{0}S_{m_{0}}(z_{0})\right)
\right)
\]%
\begin{equation}
\times\exp\left(  -i\mu\pi/2+i\sigma\phi_{b}/2+i\beta\pi/4\right)
,\label{a-mode}%
\end{equation}
with
\begin{equation}
\beta\equiv-\sigma\text{sgn}(\partial I/\partial z_{0})\text{sgn}(\partial
z/\partial z_{0}),\;\;\beta_{0}=\text{sgn}(p_{0}).\label{b-mode}%
\end{equation}

An important cautionary remark should be made. When using the stationary phase
technique we assume that the integration in Eq. (\ref{a-m-s}) goes over an
interval exceeding%
\begin{equation}
\delta z=\sqrt{\frac{2\pi}{k\left\vert \frac{\partial^{2}\Phi}{\partial z^{2}%
}\right\vert }}. \label{del-z-st}%
\end{equation}
Therefore our approach requires that%
\begin{equation}
\delta z<<\Delta z, \label{dz-Dz}%
\end{equation}
where $\Delta z$ is a difference between mode turning points. Denote by
$\Delta I$ a spread of the ray action at the range of observation. A
corresponding spread of the \textquotedblleft energy\textquotedblright\ $H$ in
accord with Eqs. (\ref{omega}) and (\ref{om-H-1}) is $\Delta H=2\pi\Delta
I/D$. Assuming that the spread of ray depths at the range of observation has
the same order of magnitude as $\Delta z$ and using Eq. (\ref{d2Phi}) we
obtain an order-of-magnitude estimate
\begin{equation}
\frac{\partial^{2}\Phi}{\partial z^{2}}\approx\frac{1}{p}\frac{\Delta
H}{\Delta z}=\frac{1}{p}\frac{2\pi}{D}\frac{\Delta I}{\Delta z},
\label{Phi-zz-approx}%
\end{equation}
where $p$ is an rms value of momentum at range $r$. As it follows from Eq.
(\ref{stationary}) an initially excited mode breaks up into a group of
approximately%
\begin{equation}
\Delta m=k\Delta I \label{Dm}%
\end{equation}
modes. Using a rough estimate $pD/2\approx\Delta z$ and combining Eqs.
(\ref{del-z-st}), (\ref{Phi-zz-approx}), and (\ref{Dm}) we get%
\begin{equation}
\delta z\approx\frac{\Delta z}{\sqrt{\Delta m/2}}. \label{del-z-st1}%
\end{equation}
The condition (\ref{dz-Dz}) is met only if $\Delta m>>1$, that is if many
modes are excited. Thus, Eq. (\ref{a-mode}) is valid only in a strongly
nonadiabatic environment and at long enough ranges where an initially excited
mode breaks up into many modes. In contrast, a point source usually excites a
large number of modes and therefore the condition (\ref{dz-Dz}) practically
does not restrict applicability of Eq. (\ref{a-s-I}).

\section{Variations of action variable and mode coupling
\label{sec:variations}}

\subsection{Number of interacting modes and applicability of adiabatic
approximation \label{sub:criterium}}

\bigskip First, consider the case when only one mode is excited at $r=0$. In
spite of the remark made at the end of Sec. \ref{sub:two-sources} the
condition (\ref{stationary}) even at short ranges properly indicates rays
contributing to the given mode. Therefore Eq. (\ref{Dm}) provides an estimate
for the number of interacting modes valid at any range. Correspondingly, the
applicability condition of adiabatic approximation can be formulated in the
form%
\begin{equation}
\Delta I<<1/k.\label{validity1}%
\end{equation}
The quantity $\Delta I$ on the left represents a difference between initial
and final actions along a typical ray. Substituting Eq. (\ref{dGdr}) into the
first of ray equations (\ref{dIdr-H-1}) yields%
\[
\Delta I=I(r)-I_{m_{0}}%
\]
\begin{equation}
=-\frac{1}{\omega}\sum_{\nu=1}^{\infty}\int_{0}^{r}dr~V_{\nu}(I_{m_{0}}%
,r)\cos\left(  \nu\theta\right)  .\label{dIdr-explicit}%
\end{equation}
In the integrands on the right we neglect the deviation of action from its
starting value. In the same approximation the second of ray equations
(\ref{dIdr-H-1}) gives%
\begin{equation}
\theta=\theta_{0}+\int_{0}^{r}\omega(I_{m_{0}},r)dr.\label{T-adiab}%
\end{equation}
Substituting this into Eq. (\ref{dIdr-explicit}) we see that Eq.
(\ref{validity1}) agrees with the condition (\ref{qm-small}).

Since we consider a linear problem and the wave field can be always
represented as a superposition of normal modes the criterion (\ref{validity1})
remains valid for an arbitrary source. But in the case of a point source we
have an explicit expression for the mode amplitude (Eq. (\ref{a-s-I})) which
can be used at both short ranges where the adiabatic approximation is still
valid and at long ranges where this approximation fails. Proceeding from this
result we shall derive a more accurate criterion. It is natural to expect that
the deviation from the adiabaticity first reveals itself in a deviation of the
phase $k\Phi$ defined by Eqs. (\ref{Phi}), (\ref{stationary}), and
(\ref{sigma}) from its value obtained under assumption that $\Lambda$ can be dropped.

In order to estimate this phase deviation take a ray (we shall call it
nonadiabatic) contributing to the $m$-th mode and compare its phase with that
of a similar ray (adiabatic) whose trajectory satisfies Eqs. (\ref{dIdr-H-1})
with $\Lambda=0$. Both rays escape the same point source and at range $r$ have
the same value of action $I=I_{m}$. Formally, we shall assume that $\Lambda$
is proportional to some small parameter $\varepsilon$. Our purpose is to
evaluate the difference between phases of the rays up to terms of order
$O(\varepsilon)$. We shall use the symbols $\delta S$, $\delta I$,
$\delta\theta$, and $\delta z$ to denote the difference in eikonals, action
variables, angle variables, and vertical coordinates of the nonadiabatic and
adiabatic rays, respectively.

Compare eikonals (see Eq. (\ref{S-I-T})) of our rays. From Eqs. (\ref{canon1})
it follows that
\begin{equation}
G(z_{0},I_{m}+\delta I_{0},0)-G(z_{0},I_{m},0)=\theta(0)\delta
I(0),\label{dle-G0}%
\end{equation}
and%
\begin{equation}
G(z+\delta z(r),I_{m},r)-G(z,I_{m},r)=p_{m}\delta z(r).\label{del-G1}%
\end{equation}
Making use of ray equations (\ref{dIdr-H-1}) yields%
\[
\delta\int_{r_{0}}^{r_{1}}\left(  Id\theta-Hdr\right)
\]
\[
=\int_{r_{0}}^{r_{1}}\left(  I_{m}d\delta\theta+\delta Id\theta-\frac{\partial
H}{\partial I}\delta Idr\right)
\]%
\begin{equation}
=I_{m}\left(  \delta\theta(r)-\delta\theta(0)\right)  .\label{del-S0}%
\end{equation}
Combining Eqs. (\ref{del-G1}) and (\ref{del-S0}) we find
\begin{equation}
\delta S=p_{m}\delta z(r)-\int_{r_{0}}^{r_{1}}\Lambda dr.\label{del-S}%
\end{equation}
The integration goes along the adiabatic ray. We assume that the
nonadiabaticity is so week that the constants $\mu$, $\beta$, and $\sigma$ for
both rays are the same. According to Eq. (\ref{Sm}) the difference between
$S_{m}$ corresponding to our rays can be estimated as $\delta S_{m}%
=p_{m}\delta z(r)$. \ Taking into account Eq. (\ref{sigma}) we finally arrive
at
\begin{equation}
\delta\Phi=-k\int\Lambda dr.\label{del-Phi}%
\end{equation}
A slightly different derivation of Eq. (\ref{del-Phi}) is given in Ref.
\cite{V2004}. Thus, applicability of adiabatic approximation requires that%
\begin{equation}
\left\vert k\int\Lambda dr\right\vert \ll\pi.\label{validity}%
\end{equation}
Using an explicit expression for $\Lambda$ given by Eq. (\ref{dGdr}) and
approximating $\theta$ by expression presented in Eq. (\ref{T-adiab}) it is
not difficult to show that the condition (\ref{validity}) agrees with Eqs.
(\ref{qm-P}) and (\ref{qm-small}).

The criterion (\ref{validity1}) is much more convenient for practical
applications than (\ref{qm-small}) and (\ref{validity}). Indeed, evaluation of
$\Delta I$ can be performed using a standard ray code without exploiting
formula (\ref{dIdr-explicit}). To find action $I$ at the given range $r$ one
should (i) compute the ray parameters $p$ and $z$ at this range and (ii) using
them as initial conditions evaluate (with the same ray code) integral
(\ref{I-P}) over the ray cycle in a corresponding reference waveguide.

\subsection{Examples \label{sec:numeric}}

To verify and illustrate the above results we have computed wave fields in two
range-dependent hydroacoustic waveguides. This is done using the code MMPE
\cite{MMPE} originally created for solving the wide angle parabolic equation.
It has been slightly modified to use the standard parabolic equation
approximation. All numerical results presented in this section have been
obtained for monochromatic wave fields at a carrier frequency of 200 Hz.

In our first example the sound speed field is taken in the form $c(r,z)=\bar
{c}(z)+\delta c(r,z)$. A range independent constituent%

\begin{equation}
\bar{c}(z)=c_{0}\left(  1+\varepsilon\left(  e^{2(z-z_{a})/B}-2(z-z_{a}%
)/B-1\right)  \right)  , \label{Munk}%
\end{equation}
with $c_{0}=1.5$ km/s, $B=1$ km, $z_{a}=-1$ km, and $\varepsilon=0.0057$
represents the so-called Munk profile widely used to study sound propagation
in deep sea \cite{BL91,JKPS94}. We consider a strong range-dependent
perturbation modelling a synoptic eddy%
\begin{equation}
\delta c(r,z)=c_{2}\,\exp\left(  -\frac{(r-r_{2})^{2}}{\Delta r^{2}}%
-\frac{(z-z_{2})^{2}}{\Delta z(r)^{2}}\right)  , \label{ring}%
\end{equation}
where
\begin{equation}
\Delta z(r)=\Delta z_{c}-\Delta z_{v}\exp\left(  -\frac{(r-r_{v})^{2}}{\Delta
r_{v}}\right)  . \label{del-z}%
\end{equation}
The following values of parameters have been selected: $c_{2}=-0.01$ km/s,
$r_{2}=300$ km, $z_{2}=-1$ km, $\Delta r=80$ km, $\Delta z_{c}=0.5$ km,
$\Delta z_{v}=0.25$ km, $r_{v}=320$ km, $\Delta r_{v}=20$ km. The isolines of
the total sound field are shown in Fig. 1.

\begin{figure}[h]
\begin{center}
\includegraphics[
height=4.8193cm, width=6.6cm ]{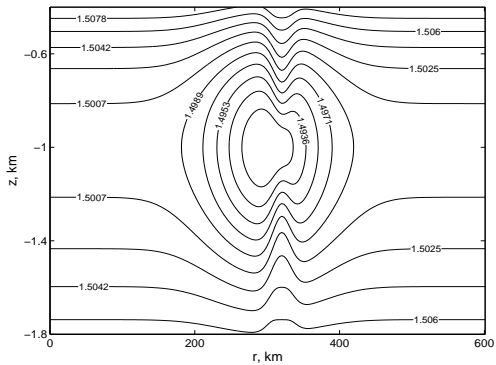}
\end{center}
\caption{Sound speed field modelling a synoptic eddy in a deep
sea. Contour
labels are of sound speed in km/s.}%
\end{figure}

The upper panel in Fig. 2 shows the deviation of mode amplitude at range
$r=600$ km from its starting value at $r=0$. The deviation is taken relative
to the rms mode amplitude at $r=0$. It is seen that only modes with $m<15$,
whose amplitudes remain practically unchanged, are adiabatic. This fact agrees
with predictions following from Eqs. (\ref{validity1}) and (\ref{validity}%
).The middle and lower panels in Fig. 2 present values of $\delta\Phi$ and
$k\Delta I=k(I(r)-I_{m})$ computed along ray paths escaping the point source
with starting values of action variables equal to $I_{m}$ ($m=0,\ldots,100$).
For each mode there are two such rays with launch angles equal in absolute
value and opposite in sign (see comment after Eq. (\ref{sigma})). Solid
(dashed) curves in both panels correspond to rays starting upward (downward).
Consistent with our expectation, although $\delta\Phi/(2\pi)$ and $k(I-I_{m})$
are not close they have the same order of magnitude and both are very small
for adiabatic modes.

\begin{figure}[h]
\begin{center}
\includegraphics[
height=4.8193cm, width=6.6cm ]{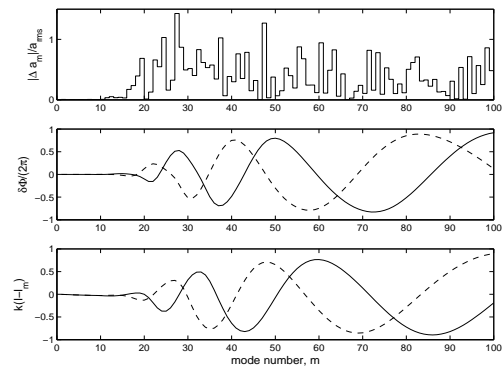}
\end{center}
\caption{Modes excited by a point source at a carrier frequency
200 Hz. The source is set at $z=-1$ km. Upper panel: Normalized
differences between mode amplitudes at ranges $r=0$ and $r=600$
km; $\Delta a_{m}=a_{m}(600$ km$)-a_{m}(0)$, $a_{\text{rms}}$ is a
rms mode amplitude at $r=0$. Middle panel: phase variations
$\delta\Phi$ due to nonadiabaticity predicted by Eq.
(\ref{del-Phi}) at $r=600$ km for rays starting upward (solid
line) and downward (dashed line). Lower panel: deviations of ray
actions from their starting values $I_{m}$ at $r=600$ km computed
for rays starting upward (solid
line) and downward (dashed line).}%
\end{figure}

The upper panel in Fig. 3 shows the range dependences of mode
amplitudes in a situation when a single mode with $m=60$ is
excited at $r=0$. Due to scattering at the eddy the $60$-th mode
breaks up into a group of modes. In order to predict the width of
this group using Eq. (\ref{Dm}) we have traced a fan of $40$ rays
with initial actions equal to $I_{60}$. The rays start from $20$
points uniformly sampling the depth interval between the turning
points of mode 60. There are two rays with starting momenta
defined by Eq. (\ref{p0}) escaping each point. The lower panel of
Fig. 3 presents the deviation of action from its starting value as
a function of range. It is clearly seen that in agreement with Eq.
(\ref{Dm}) $k\Delta I$ with $\Delta I=\max\left\vert
I(r)-I_{60}\right\vert $ representing the spread of actions at the
given range, properly predicts the width of the group of excited
modes.

\begin{figure}[h]
\begin{center}
\includegraphics[
height=6.8193cm, width=6.6cm ]{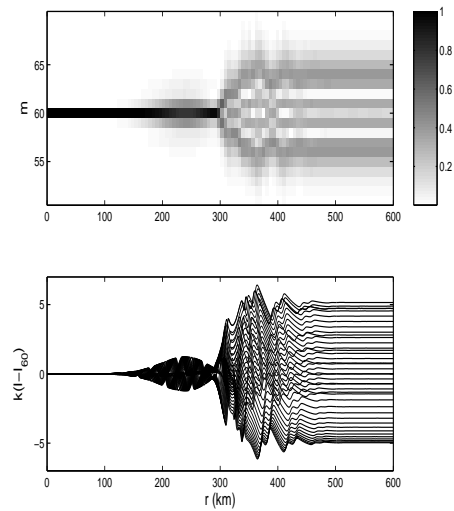}
\end{center}
\caption{Upper panel: range dependences of mode amplitudes (shown
by gradations of gray) for the case when only one mode with $m=60$
is excited at $r=0$. Lower panel: deviation of the action variable
from its starting value computed for a fan of 40 rays starting
from different depth points with
actions $I(0)=$ $I_{60}$.}%
\end{figure}

In another example of the range-dependent waveguide the sound speed field
\begin{equation}
c(r,z)=\frac{R-r}{R}c_{1}(z)+\frac{r}{R}c_{2}(z), \label{crz-lin}%
\end{equation}
with $R=200$ km being the waveguide length, represents a linear superposition
of two profiles $c_{1}(z)$ and $c_{2}(z)$. An evolution of the sound speed
profile with range is shown in Fig. 4. In this waveguide the mode coupling is
rather strong and values of $\delta\Phi$ and $k\Delta I$ (not shown) at range
200 km are on the order of unity for all modes. To check the applicability of
estimate (\ref{Dm}) we, once again, have considered the case when only the
$60$-th mode is excited at $r=0$. Figure 5 again demonstrates that $k\Delta I$
provides a good estimate for the width of a group of normal modes into which
the original mode breaks up. A new feature of function $\Delta I(r)=\max
\left\vert I(r)-I_{60}\right\vert $ absent in Fig. 3 is an appearance of
narrow spots in both panels at ranges of about 47, 94, 141, and 188 km. From
the viewpoint of the ray-based approach this phenomenon is related to the fact
that the sound speed field (\ref{crz-lin}) is a linear function of range.
Since $U(r,z)\simeq(c(r,z)-c_{0})/c_{0}$ is an almost linear function of $r$,
the derivative $\partial U/\partial r$ is practically range-independent and
the same is true of the coefficients $V_{\nu}$ present in Eq.
(\ref{dIdr-explicit}). At not very long ranges $\theta$ $\approx$ $\theta
_{0}+\omega(I_{60},0)r$ and $\left\vert I-I_{60}\right\vert $ in this
approximation vanishes at ranges equal to integer multiples of the cycle
length of the $60$-th mode. The latter is about $47$ km.

\begin{figure}[h]
\begin{center}
\includegraphics[
height=4.8193cm, width=6.6cm ]{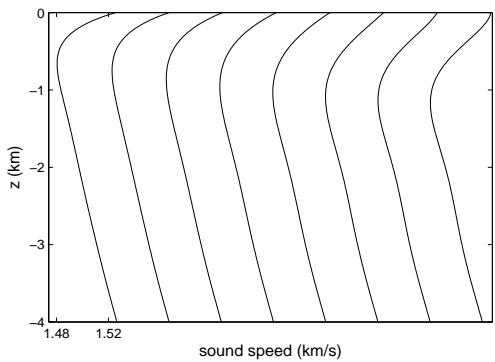}
\end{center}
\caption{Sound speed profiles at range points equally spaced
within an interval from $r=0$ to $r=200$ km are plotted (from left
to right) with a sound speed offset of $0.4$ km/s. The leftmost
and rightmost curves represent
functions $c_{1}(z)$ and $c_{2}(z)$ in Eq. (\ref{crz-lin}), respectively.}%
\end{figure}

\begin{figure}[h]
\begin{center}
\includegraphics[
height=4.8193cm, width=6.6cm ]{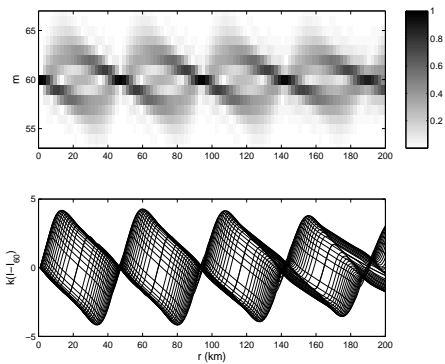}
\end{center}
\caption{The same as in Fig. 3 but constructed for a waveguide
with the sound
speed field defined by Eq. (\ref{crz-lin}) and shown in Fig. 4.}%
\end{figure}

\section{Conclusion \label{sec:concl}}

In this paper we have considered a ray-based method representing an analog of
the geometrical optics for modes. In the scope of this approach the mode
amplitude is expressed through solutions of the ray equations. It turns out
that the $m$-th mode is formed by contributions from rays whose action
variables at the range of observation, up to a multiplicative constant, are
equal to $m$. Since the ray method remains valid in the nonadiabatic
waveguide, the relatively simple formulas connecting rays and modes provide a
convenient tool for studying the applicability of adiabatic approximation. In
the present paper we have used this option.

Our simplest criterion is given by Eq. (\ref{validity1}) that imposes
limitations on the range variation of the action variable. This equation
establishes connection between the validity of adiabatic approximations for
rays and modes. It is interesting that when the condition (\ref{validity1})
fails, the quantity $k\Delta I$ estimates a number of interacting modes. For
the case of a point source we have derived a more accurate criterion
(\ref{validity}) whose implementation, however, requires more detailed
calculations. Nevertheless, it should be emphasized that (i) the left hand
sides of both Eqs. (\ref{validity1}) and (\ref{validity}) can be evaluated
using a standard ray tracing and (ii) both criteria agree with the condition
(\ref{qm-small}) derived from the mode coupling equation. Note also that
unlike conditions (\ref{DL}) and (\ref{k2DL}), our criteria account for
accumulation of errors with range.

Since the left hand sides of Eqs. (\ref{validity1}) and (\ref{validity}) are
proportional to the carrier frequency it is clear that the lower the
frequency, the better these inequalities are satisfied and, hence, the wider
the validity region of the adiabatic approximation. The same conclusion
follows from Eq. (\ref{k2DL}).

Finally, notice that all our results have been obtained in the high frequency
approximation which, naturally, restricts their generality. In contrast, the
use of the parabolic equation approximation is not principal. The results can
be easily generalized to the case when the wave field is governed by the
Helmholtz equation. Although we have considered sound waves propagating in
underwater acoustic waveguides the formulas derived here can be applied for
description of wave propagation in different waveguide media and for analysis
of a quantum particle oscillating in a potential well with time-dependent parameters.

This work was supported by the Russian Foundation for Basic Research under
Grant No. 03-02-17246.

\end{document}